\documentclass[aps,prb,reprint,showpacs,superscriptaddress]{revtex4-1}
\usepackage{graphicx}
\usepackage{amsmath}
\usepackage{amssymb}
\usepackage{dcolumn}
\usepackage{dsfont}
\usepackage{latexsym}
\usepackage{rotating}
\usepackage{color}
\usepackage{latexsym}
\usepackage{bbm}
\usepackage{subfigure}
\usepackage{float}
\usepackage{epsfig}
\usepackage{psfrag}
\usepackage{natbib}
\usepackage{bm}
\usepackage{amsthm}
\usepackage{eucal}
\usepackage{mathrsfs}
\usepackage{color} 

\usepackage{hyperref}
\hypersetup{
colorlinks=true,final=true,
        linkcolor=blue,
        citecolor=blue,
        filecolor=blue,
        urlcolor=blue,
}

\begin{document}

\title{Phase segregation of superconductivity and ferromagnetism at LaAlO$_3$/SrTiO$_3$ interface}

\author{N. Mohanta} \email{nmohanta@phy.iitkgp.ernet.in}
\affiliation{Department of Physics, Indian Institute of
  Technology Kharagpur, W.B. 721302, India}
\author{A. Taraphder} \email{arghya@phy.iitkgp.ernet.in} 
\affiliation{Department of Physics, Indian Institute of
  Technology Kharagpur, W.B. 721302, India}
\affiliation{Centre for Theoretical Studies, Indian Institute of
  Technology Kharagpur, W.B. 721302, India}

\begin{abstract}
The  highly  conductive two-dimensional  electron  gas  formed at  the
interface   between   insulating   SrTiO$_3$   and   LaAlO$_3$   shows
low-temperature   superconductivity   coexisting  with   inhomogeneous
ferromagnetism. The Rashba spin-orbit interaction with in-plane Zeeman
field of  the system favors  $p_x \pm ip_y$-wave  superconductivity at
finite momentum. Owing to the intrinsic disorder at the interface, the
role of spatial inhomogeneity on the superconducting and ferromagnetic
states becomes important. We find that for strong disorder, the system
breaks  up into  mutually  excluded regions  of superconductivity  and
ferromagnetism. This  inhomogeneity-driven electronic phase separation
accounts  for   the  unusual  coexistence   of  superconductivity  and
ferromagnetism observed at the interface.
\end{abstract}


\maketitle

\section{Introduction}
The  physics underlying  the interface  sandwiched  between insulating
oxides SrTiO$_3$ and LaAlO$_3$  has generated tremendous excitement in
the  last  few  years after  the  discovery  of  a high  mobility  two
dimensional  electron gas  (2DEG) at  the interface~\cite{Ohtomo2004}.
In 2007, Reyren,  \textit{et al.}~\cite{Reyren31082007} found that the
system is superconducting below 200 mK. The other intriguing phenomena
the   2DEG  exhibits  are   electric  field   induced  metal-insulator
transition~\cite{Cen2008,Caviglia2008,lu:172103}                    and
superconductor-insulator transition~\cite{PhysRevLett.103.226802}.  In
particular,  there  are strong  evidences  that the  superconductivity
coexists  with a finite magnetic moment  (0.3-0.4~$\mu_B$  per interface
unit  cell)~\cite{PhysRevLett.107.056802}. Torque  magnetometry  and
transport measurements report an  in-plane magnetic ordering from well
below the superconducting  transition to 200 K~\cite{Li2011}. Scanning
squid data reveals that, the sample contains sub-micrometer patches of
ferromagnetic   domain  within  a   non-uniform  background   of  weak
diamagnetic superconducting susceptibility~\cite{Bert2011}.

The highly unusual coexistence of superconductivity and ferromagnetism
has  spawned several  theoretical  efforts to  unravel  the nature  of
states  at   the  interface.  As  proposed   by  Pavlenko,  \textit{et
  al.}~\cite{PhysRevB.85.020407},  magnetism may  not be  an intrinsic
property of  the 2DEG, but  is a result  of the spin splitting  of the
populated  electronic  states  induced  by  Oxygen  vacancies  in  the
SrTiO$_3$ or  LaAlO$_3$ layer. The metallic behavior  of the interface
results  from  the  2D  electron  liquid produced  by  the  electronic
reconstruction.  The metallic  state is  related to  a superconducting
state  below  200  mK,   occurring  in  regions  of  vanishing  Oxygen
vacancies.   Michaeli,  \textit{et  al.}~\cite{PhysRevLett.108.117003}
argued that due to strong spin-orbit interaction superconductivity and
ferromagnetism can coexist via Fulde-Ferrell-Larkin-Ovchinnikov (FFLO)
type pairing at finite momentum;  the FFLO state being quite sensitive
to disorder while possibly even stabilized by strong disorder.

According  to  Fidkowski,  \textit{et  al.}~\cite{PhysRevB.87.014436},
droplets  of  superconductivity  are  formed in  the  bulk  insulating
SrTiO$_3$ due  to some  imperfections in the  lattice. In  presence of
LaAlO$_3$, the mobile electrons in the 2DEG mediate a coupling between
the   droplets   which  percolate   and   grow   until  a   long-range
superconducting        order       forms.         Another       recent
suggestion~\cite{2013arXiv1304.2970C}  is multi-band superconductivity
resulting    from   percolation    of   filamentary    structures   of
superconducting     puddles.     Recently,     Randeria,    \textit{et
  al.}~\cite{Banerjee2013}    proposed   that    the    ground   state
magnetization  is   a  long-wavelength  spiral  aligning   to  a  weak
ferromagnetism  in the  presence of  applied magnetic  field.  Several
studies have  provided hints, but  not yet conclusive support  for the
existence   of   superconductivity   and  ferromagnetism   and   their
coexistence.

It  is  now  well-understood  that  the  metallic  conduction  at  the
interface is  mainly due to two  reasons: the Oxygen  vacancies at the
interface~\cite{PhysRevX.3.021010,  PhysRevB.86.064431,Schlom2011} and
an intrinsic electronic transfer  mechanism known as polar catastrophe
in which  half an  electronic charge is  transferred to  the interface
~\cite{Nakagawa2006,PhysRevB.80.075110}.   The    electrons   at   the
interface, which is TiO$_2$ terminated, occupy the 3d $t_{2g}$ orbital
of Ti  atoms.  Michaeli, \textit{et al.}~\cite{PhysRevLett.108.117003}
pointed  out   that  there   are  three  different   bands,  uniformly
distributed  at the interface,  responsible for  superconductivity and
ferromagnetism. The $d_{xy}$ band is wider than the $d_{xz}$, $d_{yz}$
bands and is  relatively lower in energy at the  $\Gamma$ point. It is
therefore likely that the electrons  in this band get localized at the
interface  sites  due  to  Coulomb  correlation  and  eventually  form
localized moments. A  ferromagnetic local exchange interaction between
the conduction  band electrons and  the local moments  will thereafter
lead to an effective in-plane Zeeman field. The Zener kinetic exchange
may  then order local  moments as  well. Very  recently, spectroscopic
studies  provided  direct  evidence  for  in-plane  ferromagnetism  of
electrons  with $d_{xy}$  character~\cite{Lee2013}.  Another important
feature of  the system is  the large Rashba coupling  which originates
from the  broken inversion  symmetry in the  two-dimensional interface
plane.

It is more  or less evident that inhomogeneities  at the interface, in
the  form   of  Oxygen  vacancies  and  intrinsic   disorder,  are  an
inseparable part  of the interface.  It is therefore likely  that they
have a profound effect on  the long range orders and their coexistence
at  the interface.  In the  following analysis,  we  use inhomogeneous
Bogoliubov-de Gennes (BdG) theory to study the effects of non-magnetic
disorder  on  superconductivity  and  magnetism at  the  interface  in
presence of a strong Rashba  coupling.  Our calculations show that due
to large  disorder, the system  forms superconducting islands.  At the
regions  where superconductivity is  destroyed by  disorder, electrons
order  ferromagnetically along  the plane  due to  the  in-plane field
forming ferromagnetic domains.   This disorder-driven electronic phase
separation enables superconductivity  and ferromagnetism to coexist in
the same sample  in electronically phase-separated regions. Electronic
phase separation as a  mechanism for coexistence of various long-range
orders   has   been   established    in   manganites   in   the   last
decade~\cite{PhysRevLett.80.845,Dagotto20011}.    Its   role  in   the
interfaces has  also been  posited by some  experimental groups  and a
theoretical  analysis in  the present  context is  therefore eminently
topical.

The rest of the paper is organized as follows. In the next section, we
describe our  model for a Rashba  spin-orbit-coupled superconductor in
presence of an  in-plane Zeeman field. In section~\ref{bdg_treatment},
starting with the standard  inhomogeneous BdG mean-field formalism, we
present  and   analyze  the   effects  of  spatial   inhomogeneity  on
superconductivity and ferromagnetism and their possible coexistence at
the  interface.  In  addition,   we  extend  our  analysis  to  finite
temperatures and study the coexistence. Sections~\ref{discussions} and
~\ref{conclusion}  are  for  discussions  of  results  and  concluding
remarks.

\vspace{-1.9em}
\section{Model}
\label{model}
\begin{figure}[!ht]
\begin{center}
\epsfig{file=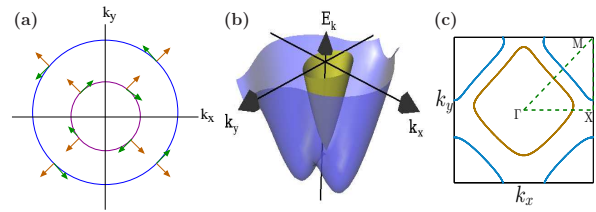, width=80mm}
\caption{(Color   online)  (a)  Two-sheeted   Fermi  surface   of  the
  Rashba-splited  bands showing  the  helical alignment  of the  spins
  (green)  and  corresponding momentum  (orange).   The (b)  resultant
  bands and (c) asymmetric  Fermi surface created by Rashba spin-orbit
  interaction with Zeeman field along $\hat{x}$ direction.}
\label{bands_fs}
\end{center}
\end{figure}

\vspace{-0.5em}
Electrons at  the interface occupy the  $t_{2g}$ bands of  Ti atom and
therefore   these   bands   are   thought  to   be   responsible   for
superconductivity  and  ferromagnetism in  the  system. The  electrons
occupying $d_{xy}$ band, which is much wider than $d_{xz}$ or $d_{yz}$
bands and is lower in energy  at the $\Gamma$ point, are localized due
primarily to  the electron-electron correlation at  the interface. The
local moments interact via  exchange interaction with the electrons in
conduction band (claimed to  be slightly below the terminating TiO$_2$
layer in the  SrTiO$_3$ side~\cite{PhysRevLett.108.117003}) leading to
an  in-plane  Zeeman   field.   Superconductivity  originates  from  a
short-range electron-electron  attractive interaction of  strength $U$
(possibly retarded by the  phonon energies). The following Hamiltonian
describes a  Rashba spin-orbit  coupled superconductor in  an in-plane
Zeeman field.
\begin{equation}
\begin{split}
&{\cal H}={\cal H}_0+{\cal H}_{ex}+{\cal H}_{so}+{\cal H}_{sc},\\
&{\cal H}_0=\sum_{k,\sigma}\epsilon_k c_{k\sigma}^\dagger c_{k\sigma}\\
&{\cal H}_{ex}=-\sum_{k,\sigma,\sigma^{\prime}}( \boldsymbol{H} \cdot \boldsymbol{\sigma})_{\sigma,\sigma^{\prime}}c_{k\sigma}^\dagger c_{k\sigma^{\prime}}\\
&{\cal H}_{so}=\alpha \sum_{k,\sigma,\sigma^{\prime}} ( \boldsymbol{R}(\boldsymbol{k}) \cdot \boldsymbol{\sigma})_{\sigma,\sigma^{\prime}} c_{k\sigma}^\dagger c_{k\sigma^{\prime}}\\
&{\cal H}_{sc}=-U\sum_{k,k'} c_{k\uparrow}^\dagger c_{-k\downarrow}^\dagger c_{k'\downarrow} c_{-k'\uparrow}
\end{split}
\label{hamiltonian}
\end{equation}
\noindent  where,  $\epsilon_k =-2t(\cos  k_x  +  \cos k_y)-\mu$  with
hopping  amplitude $t$  in  a square  lattice  and chemical  potential
$\mu$. The in-plane  Zeeman field is given by  $H=(H_x,H_y,0)$ and the
Rashba  coupling  is   $R({\boldsymbol{k}})=(\sin  k_y,  -\sin  k_x)$,
$\boldsymbol{\sigma}$ being the Pauli matrices.
The Rashba  spin-orbit interaction creates helical bands  in which the
electron  spins   are  aligned  with  respect  to   the  direction  of
propagation  as shown  in Fig.~\ref{bands_fs}(a).   The  energy bands,
created by  the Rashba  SOC with in-plane  Zeeman field, are  given by
\begin{equation}
\epsilon_{\pm}(\mathbf{k})=\epsilon_k \pm |\alpha \boldsymbol{R}(\boldsymbol{k})-\boldsymbol{H}|
\end{equation}
\noindent  and the corresponding
eigenstates $c_{k,\pm}$ are obtained by the following transformation
\begin{equation}
\begin{split}
\begin{pmatrix} \begin{array}{c} c_{k\uparrow} \\ c_{k\downarrow} \end{array} \end{pmatrix} &= \frac{1}{\sqrt{2}}\begin{pmatrix} \begin{array}{cc} 1 & 1\\e^{i\phi_{k}} & -e^{i\phi_{k}} \end{array} \end{pmatrix} \begin{pmatrix} \begin{array}{c} c_{k,+} \\ c_{k,-} \end{array} \end{pmatrix}
\end{split}
\end{equation}
\noindent where $\phi_{k}=\tan^{-1}(\frac{-\sin k_x-H_y}{\sin k_y-H_x})$.

\noindent In the chiral  basis $[c_{k,+}, c_{k,-}]$, ${\cal H}_0+{\cal
  H}_{ex}+{\cal  H}_{so}$  is diagonal  and  ${\cal  H}_{sc}$ has  the
pairing symmetry of $p_x \pm ip_y$ superconductivity:
\begin{equation}
{\cal H}_{sc}=\sum_k \Delta(e^{-i\phi_{k}}c_{k,+}^{\dagger}c_{-k,+}^{\dagger}+e^{i\phi_{k}}c_{k,-}^{\dagger}c_{-k,-}^{\dagger}+h.c.)
\end{equation} 
\noindent with $\Delta$ as the  pairing amplitude. The Zeeman field in
the in-plane direction shifts the Fermi surface from the center of the
Brillouin  zone,  as  shown  in Fig.~\ref{bands_fs}(b)  and  (c),  and
therefore removes the  degeneracy of a pair of  electrons on the Fermi
surface   with   opposite    wave   vectors   ${\boldsymbol{k}}$   and
$-{\boldsymbol{k}}$. Hence for pairing  of electrons with same energy,
it turns out to be energetically  favorable to have a finite center of
mass       momentum      (proportional      to       the      in-plane
field~\cite{PhysRevB.81.184502}) of the pair.

In a clean system, the field due to moments at the interface is fairly
strong  and   may  even   be  larger  than   superconducting  critical
field~\cite{PhysRevLett.108.117003}. In  that case the  system becomes
magntically ordered.  On the  other hand, if  the field is  not strong
enough,    supreconductivity   may    dominate   in    a   homogeneous
system.   However,  disorder   has   a  significant   effect  on   the
superconducting state (particularly for  non-s-wave cases) and in turn
lead to  islands of non-superconducting regions.  Phase segregation in
real space  is a likely  outcome in such inhomogeneous  situations and
superconductivity   and    ferromagnetism   coexist.   The   following
Bogoliubov-de  Gennes analysis explores  the role  of disorder  on the
superconductivity in such a system.

\section{Role of spatial Inhomogeneity}
\label{bdg_treatment}
The  interplay  between  superconductivity  and disorder  has  been  a
central     issue     of     many     recent     investigations     on
superconductivity~\cite{PhysRevB.65.014501,dubinature2007,Chatterjee2008582}.
With  increasing  disorder,  there  exists  a  phase  transition  from
superconductor     to     insulator     in    two-dimensional     thin
films~\cite{PhysRevLett.65.923}.   The  Rashba spin-orbit  interaction
induces chiral $p_x  \pm ip_y$-wave superconductivity where Anderson's
theorem,  unlike in  conventional \textit{s}-wave  superconductors, is
not  applicable. Hence,  it  is  interesting to  study  the effect  of
non-magnetic   disorder  on   a  two-dimensional   system   with  both
ferromagnetism  and unconventional superconductivity,  especially when
there is broken mirror symmetry.

We  consider  the  following  BdG  Hamiltonian for  a  Rashba  coupled
superconductor  on a  square lattice  with an  effective  Zeeman field
along the $\hat{x}$ direction (without any loss of generality only the
$x$-component of the field is considered for simplicity),
\begin{equation}
\begin{split}
{\cal H}_{BdG}&=-t\sum_{<ij>,\sigma}(c_{i\sigma}^\dagger c_{j\sigma}+h.c.)-
\sum_{i,\sigma}(\mu-V{_i}) c_{i\sigma}^\dagger c_{i\sigma}\\
&-H_x\sum_{i,\sigma,\sigma^{\prime}}(\sigma_x)_{\sigma \sigma^{\prime}}c_{i\sigma}^\dagger c_{i\sigma^{\prime}}+ \sum_{i}\Delta(r_i)(c_{i\uparrow}^{\dagger}c_{i\downarrow}^{\dagger}+h.c.)\\
&-i\frac{\alpha}{2}\sum_{<ij>,\sigma,\sigma^{\prime}}c_{i\sigma}^{\dagger}(\boldsymbol{\sigma}_{\sigma \sigma^{\prime}} \times \boldsymbol{d}_{ij})_z c_{j\sigma^{\prime}}
\end{split}
\label{HBdG}
\end{equation}
\noindent where  $V_i$ is  the random disorder  in the  local chemical
potential, taken  to be  uniformly distributed between  [-W,W]. ${\cal
  H}_{BdG}$ is diagonalized  via a spin-generalized Bogoliubov-Valatin
transformation
$\hat{c}_{i\sigma}(r_i)=\sum_{i,\sigma^{\prime}}u_{n\sigma\sigma^{\prime}}(r_i)\hat{\gamma}_{n\sigma^{\prime}}+v_{n\sigma\sigma^{\prime}}^*(r_i)\hat{\gamma}^{\dagger}_{n\sigma^{\prime}}$.
This  gives  the following  equation  for  the  local order  parameter
$\Delta(r_i)=-U<c_{i\uparrow}c_{i\downarrow}>$   in   terms   of   the
Bogoliubov amplitudes $u_{n\sigma}(r_i)$ and $v_{n\sigma}(r_i)$:
\begin{equation}
\begin{split}
\Delta(r_i)=&-U\sum_{n}[u_{n\uparrow}(r_i)v^*_{n\downarrow}(r_i)(1-f(E_n))\\
&+u_{n\downarrow}(r_i)v^*_{n\uparrow}(r_i)f(E_n)]
\end{split}
\label{delsi}
\end{equation}
\noindent  where $f(x)=1/(1+e^{x/{k_BT}})$  is the  Fermi  function at
temperature  $T$.  The  in-plane magnetization  density with 
exchange field along $\hat{x}$ direction is obtained via
\begin{equation}
\begin{split}
m(r_i)&=<c_{i\uparrow}^{\dagger}c_{i\downarrow}+c_{i\downarrow}^{\dagger}c_{i\uparrow}>\\
&=\sum_{n,\sigma}u_{n\sigma}^{*}u_{n\sigma^{\prime}}f(E_n)+v_{n\sigma}u_{n\sigma^{\prime}}^{*}(1-f(E_n))
\end{split}
\end{equation}
The  orbital contribution  to the  magnetic moment  is expected  to be
fairly  small in  the inhomogeneous,  phase segregated  situation with
short   superconducting    mean-free   path   ($\xi$).     In   chiral
superconductors,  even for  homogeneous condensates,  this is  a hotly
debated issue and values of orbital moment range from $\hbar$ per pair
to $k_{F}\xi$  or even $(k_{F}\xi )^2$ (i.e.,  corresponding moment of
order $\mu_{B}$ or 10$^{-6}\mu_B$;  for recent results and comments on
this,                                                               see
Ref.~\cite{1367-2630-11-5-055063,JPSJ.65.664,Stone20082,PhysRevB.74.024408}).
Clearly, in an inhomogeneous situation  such as the one obtains in the
interface (with an intrinsically short $\xi$), the total effect coming
from  the magnetic domains  would be  much reduced.   The experimental
indicators certainly do  not subscribe to a large  contribution to the
angular momentum, though careful experiments like Kerr rotation, X-ray
circular dichroism,  $\mu$SR and resonant Raman  scattering are needed
to resolve  this.  We have,  therefore, ignored this in  the foregoing
and restricted ourselves to the spin contribution only.

The      quasi-particle     amplitudes      $u_{n\sigma}(r_i)$     and
$v_{n\sigma}(r_i)$  are  determined   by  solving  the  following  BdG
equations:
\begin{equation}
\begin{split}
{\cal H}_{BdG}\phi_n(r_i)=\epsilon_n\phi_n(r_i)
\end{split}
\label{bdg eq.}
\end{equation}
where $\phi_n=[u_{n\uparrow}(r_i),u_{n\downarrow}(r_i),v_{n\uparrow}(r_i),v_{n\downarrow}(r_i)]$. 

In what  follows, Eq.(\ref{delsi}) and Eq.(\ref{bdg  eq.})  are solved
self-consistently on  a finite,  large two dimensional  square lattice
with  periodic boundary  conditions and  finally the  mean  values are
calculated     using    $\Delta=\frac{1}{N}\sum_i\Delta_s(r_i)$    and
$m=\frac{1}{N}\sum_im(r_i)$,   over   several   realizations  of   the
disorder.  N is the total  number of sites. The results presented here
are obtained with $31\times31$ lattice sites, which is large enough to
obtain a  quantitatively satisfactory  description.  We work  in grand
canonical  ensemble with  $t=1$ and  $U=1$ (consistent with previous 
work~\cite{PhysRevB.65.014501,PhysRevLett.81.3940}) and unless specified, 
the density of  electrons is kept  at half-filling. The case of filling 
away from half is discussed later. We first  discuss our results at 
$T=0$ and later extend to finite temperatures.
\begin{figure}[!ht]
\begin{center}
\epsfig{file=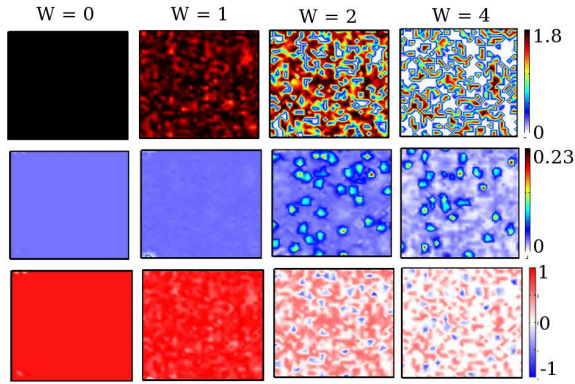, width=80mm}
\caption{(Color online) The spatial  distribution of the local pairing
  amplitude $|\Delta(r_i)|$ (top  row) and magnetization (middle row).
  The columns are for disorder strength $W  = 0$, $W = 1$, $W = 2$ and
  $W = 4$  (from left to right). The lowest  row shows the coexistence
  in combined plots of  magnetization and local pairing amplitude; red
  and    blue    represent    regions   of    superconductivity    and
  ferromagnetism. The  parameters used are  $H_x = 0.5$ and  $\alpha =
  0.8$.}
\label{disorder_maps}
\end{center}
\end{figure}

The inhomogeneous BdG method is useful to study the spatial variations
of  the  pairing  amplitude.    In  the  large  disorder  regime,  the
microscopic  details turn  out to  be  very important  leading to  new
unanticipated  results.   Fig.~\ref{disorder_maps}  shows the  spatial
distribution  of  the  local  pairing  amplitude  $|\Delta(r_i)|$  for
different  disorder  strengths.   The  top  and middle  row  show  the
development  of superconductivity and  magnetism separately  while the
bottom one reveals the full picture of coexistence.  In the homogenous
system ($W=0$),  superconductivity is  uniform at the  interface plane
and  the  magnetic response  is  very  weak.  However, in  the  highly
inhomogeneous  case ($W\geq2$),  the  disorder abets  in breaking  the
uniform suprconductivity into islands separated by non-superconducting
regions.  It is  quite  straightforward to  understand the  underlying
physical  situation  within  the  BdG  formalism.  The  regions  where
$|\mu-V_{i}|$  is small,  superconductivity sets  in easily  while the
regions  of  large number  fluctuations,  having large  $|\mu-V_{i}|$,
militate against  it. Robust ferromagnetic  puddles are formed  at the
regions where superconductivity is degraded by disorder. Thus disorder
helps the  two competing phases  to coexist by keeping  them spatially
seperated.

\subsubsection{Distribution of pairing gap and magnetization}
\begin{figure}[!ht]
\begin{center}
\epsfig{file=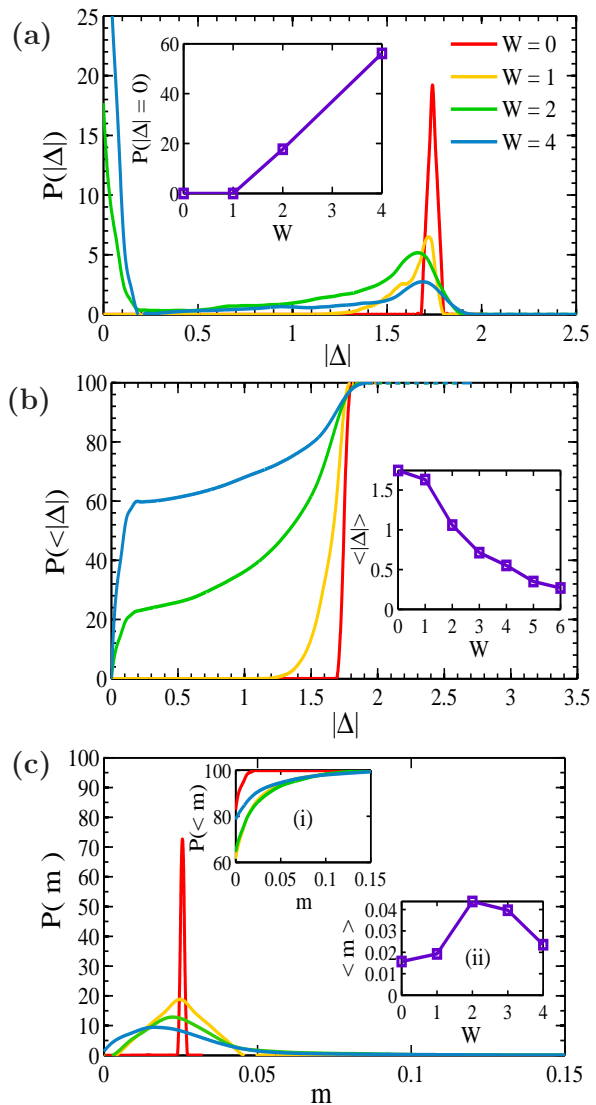, width=80mm}
\caption{(Color online) (a)  Probability distribution of local pairing
  amplitudes. Inset shows the variation of the probability at zero gap
  with  disorder strength. (b)  The curves  show the  probability P($<
  \Delta$) that the gaps are  less than a given $\Delta$ for different
  disorder strength.  Inset is $<|\Delta|>$ as a  function of disorder
  strength. (c)  The probability distribution  of local magnetization.
  Inset (i)  is P($< m$) as  a function of  $m$ and inset (ii)  is the
  variation  of average  magnetization with  disorder.  The parameters
  used are $H_x = 0.5$, $\alpha = 0.8$ at $W = 4$.}
\label{probability}
\end{center}
\end{figure}

A great deal of information about the nucleation of the pairing on the
microscopic  scale   can  be  extracted  from  the   data  plotted  in
Fig.~\ref{disorder_maps}.  In Fig.~\ref{probability}(a),  we  plot the
distribution  of the  local  pairing amplitude  $|\Delta(r_i)|$ for  a
number of disorder strengths.
\noindent  In the homogeneous  system ($W=0$),  there is  a pronounced
peak  near  $|\Delta_0|$,  the  mean-field  value at  $T=0$.   As  the
disorder  increases, the  peak  at $|\Delta_0|$  decreases, weight  is
gradually transferred to lower gap values and the probability for zero
gap  increases  (shown  in  the inset  of  Fig.~\ref{probability}(a)),
indicating  the  formation  of  non-superconducting  regions.   It  is
interesting to note that even  for low disorder strength, regions with
zero  gap appear  and  long range  superconducting  order is  severely
affected  as  expected for  a  $p_x  \pm  ip_y$ superconductivity  (in
contrast   with  the   s-wave  case   where  the   effect   is  indeed
weaker~\cite{PhysRevB.65.014501}).   From Fig.~\ref{disorder_maps}, we
extract the  probability P($<  |\Delta|$) that the  gap value  is less
than a given $|\Delta|$, as plotted in Fig.~\ref{probability}(b). This
clearly demonstrates  how the propensity towards  formation of low-gap
(and  eventually gapless)  regions increases  rapidly  with increasing
disorder.  The inset of  Fig.~\ref{probability}(b) shows that the mean
pairing  amplitude  $<|\Delta|>$  decreases strongly  with  increasing
disorder.   However, it  never vanishes,  as pointed  out  by Avishai,
\textit{et  al.}~\cite{dubinature2007}; even  in  the strong  disorder
limits there are always few superconducting regions with finite gaps.

\noindent   Disorder  has   significant  effects   on   the  resulting
ferromagnetic landscape  described by the  probability distribution of
local   magnetization   in   Fig.~\ref{probability}(c).    The   peaks
correspond  to weak  ferromagnetic background,  which  gradually shift
towards zero as disorder increases. The tail at higher magnetizations,
for  $W \ge  2$ reflects  the  formation of  ferromagnetic puddles  at
higher  disorder  strength. However,  for  very  strong disorder,  the
ferromagnetic domains  are also destroyed, reflected in  the nature of
the    variation   of   average    magnetization   (inset    (ii)   of
Fig.~\ref{probability}(c)).  The competition between superconductivity
and ferromagnetism is borne out from a comparison of the distributions
P($<  |\Delta|$)  and  P($<   m$)  for  pairing  amplitude  and  local
magnetization  respectively.  With  higher  disorder, the  probability
P$(|\Delta|)$ of gapped regions decreases (Fig.~\ref{probability}(b)),
while  the regions  of ferromagnetic  domains increase  (inset  (i) of
Fig.~\ref{probability}(c)).     Thus   the    formation    of   robust
ferromagnetism tracks the destruction of superconducting regions. Such
a correlation is crucial for the observed formation of microscopically
phase separated regions of ferromagnetism and superconductivity at the
interface~\cite{Wang2011}.

\subsubsection{Correlation functions}
To get a better understanding of the nature of superconducting regions
or  the ferromagnetic  domains and  the range  of their  order,  it is
useful   to   study   the  disorder-averaged   correlation   functions
$D_m(|r_i-r_j|)=<m(r_i)m(r_j)>$         and        $D_{sc}(|r_i-r_j|)=
<|\Delta|(r_i)|\Delta|(r_j)>$.
\begin{figure}[!ht] 
\begin{center}
\epsfig{file=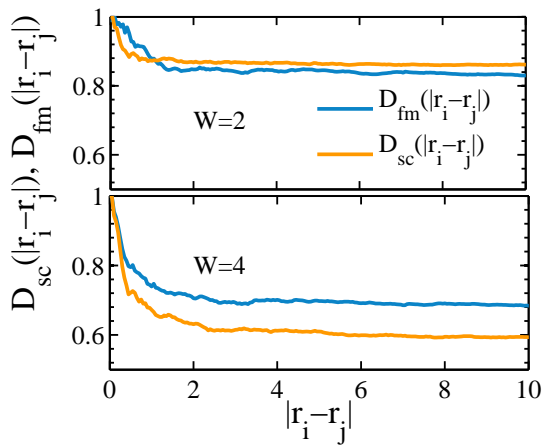, width=80mm}\vspace{-1.0em}
\caption{(Color online) Plots of the correlation 
    functions ${D_{fm}(|r_i-r_j|)}$ and ${D_{sc}(|r_i-r_j|)}$ (see text) for 
    local magnetizations and superconducting pair-amplitudes respectively as a function 
    of the separation $|r_i-r_j|$. All the plots are normalized to unity at 
    zero-separation. The parameters used are $H_x = 0.5$, $\alpha = 0.8$ at 
    $T = 0$.}
\label{corr_func}
\end{center}
\end{figure}
\noindent  As depicted in  Fig.~\ref{corr_func}, both  the correlation
functions ${D_{sc}(|r_i-r_j|)}$  and ${D_m(|r_i-r_j|)}$ falls rapidly,
indicating  the absence of  any long-range  order of  magnetization or
superconductivity.  The rapid fall of the correlation functions within
a few lattice spacings (even for a weak disorder, $W = 1$, not shown),
emphasizes the short range nature  of the underlying order. It is also
noted from  the nature  of these correlation  functions at  $W=2,\, 4$
that   superconductivity   is   more   sensitive  to   disorder   than
ferromagnetism  which persists  where  superconductivity is  destroyed
before disorder suppresses both.

\subsubsection{Local Density of States}
In a  disordered superconductor,  a very useful  quantity that  can be
seen in tunneling  experiments is the local density  of states (LDOS),
which, at zero temperature, is given by
\begin{equation}
\rho(E)=\frac{1}{N}\sum_{n,r_i,\sigma}[|u_{n\sigma}(r_{i})|^2 \delta(E-E_n)+|
v_{n\sigma}(r_{i})|^2 \delta(E+E_n)]
\end{equation}
\begin{figure}[!ht] 
\begin{center}
\epsfig{file=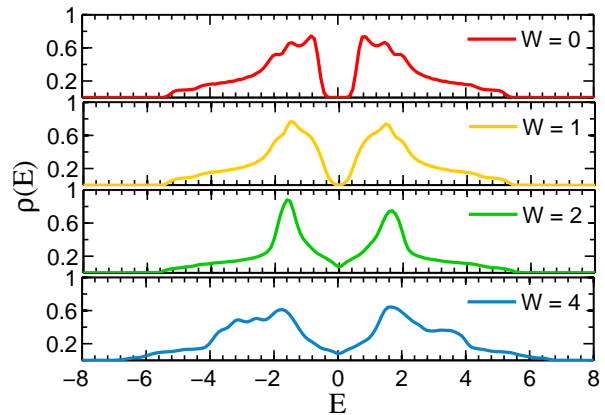, width=80mm}
\caption{(Color online) Variation of local density of states with
  disorder strength ranging from $W = 0$ (homogeneous) to $W = 4$
  (highly disordered) with $H_x = 0.5$ and $\alpha = 0.8$ at $T = 0$.}
\label{ldos}
\end{center}
\end{figure}
\noindent With increasing disorder, states  begin to appear in the gap
and   the  nature   of   tunneling  spectrum   changes  rapidly.   The
single-particle  density of  state is  plotted in  Fig.~\ref{ldos} for
different  disorder strengths. In  the homogeneous  limit, there  is a
real gap in the DOS, with the usual pile up of states on both sides of
it. With increasing disorder, a pseudo-gap appears, reminiscent of the
under-doped                                                  high-$T_c$
superconductors~\cite{0034-4885-62-1-002,PhysRevB.71.014514}    or   a
superconductor       in      presence      of       high      magnetic
field~\cite{PhysRevB.78.024502}. The  pile up also smears  out and the
spectrum  is pushed  towards  higher energy.   Therefore, in  scanning
tunneling microscopy, one would be able to observe regions of real gap
with  a  pile up  of  states  in the  DOS,  separated  by gapless  (or
pseudo-gapped) non-superconducting regions. These latter regions would
also show up clearly in a magnetic force microscopy.

\subsubsection{Superfluid Density}
To explicitly understand how disorder affects superconductivity and to
track the superconductor-insulator transition,  one needs to study the
superfluid density that  characterizes superconducting phase rigidity.
We  calculate  the  superfluid  density, defined  from  the  effective
Drude-weight~\cite{PhysRevLett.68.2830,0295-5075-34-9-705}, as
\begin{equation}
\rho_s \equiv \frac{D_s}{\pi e^2}=-<K_x>+\Pi_{xx}(q \rightarrow 0, \omega \rightarrow 0)
\end{equation}
\noindent  The first term  represents the  diamagnetic response  to an
external  magnetic field  ${\bf{B}}=\nabla \times  {\bf{A}}$  with the
local kinetic energy
\begin{equation}
\begin{split}
K_x^i=&-t\sum_{n,\sigma}(f(E_n)[u_{n\sigma}^*(r_{i+\hat{x}})u_{n\sigma}(r_{i})+c.c]\\
&+(1-f(E_n))[v_{n\sigma}^*(r_{i+\hat{x}})v_{n\sigma}(r_{i})+c.c])
\end{split}
\end{equation}
\noindent The  second term is  the paramagnetic response given  by the
disorder-averaged transverse current-current correlation function
\begin{equation}
\Pi_{xx}(q \rightarrow 0, \omega \rightarrow 0)=\frac{1}{N}\sum_{ij}\Pi_{xx}^{ij}(\omega=0)
\end{equation}
\noindent with
\begin{equation}
\begin{split}
\Pi_{xx}^{ij}&(\omega=0)\\
&=\sum_{n_1,n_2}A_{n_1,n_2}^i[A_{n_1,n_2}^{j*}+D_{n_1,n_2}^{j}]\frac{f(E_{n_1})-f(E_{n_2})}{E_{n_1}-E_{n_2}}
\end{split}
\end{equation}
where 
\begin{equation}
\begin{split}
&A_{n1,n2}^i=2[u_{n_1}^*(r_{i+\hat{x}})u_{n_2}(r_i)-u_{n_1}^*(r_i)u_{n_2}(r_{i+\hat{x}})]\\
&D_{n1,n2}^i=2[v_{n_1}(r_{i+\hat{x}})v_{n_2}(r_i)^*-v_{n_1}(r_i)v_{n_2}^*(r_{i+\hat{x}})]\\
\end{split}
\end{equation}
\begin{figure}[!ht] 
\begin{center}
\epsfig{file=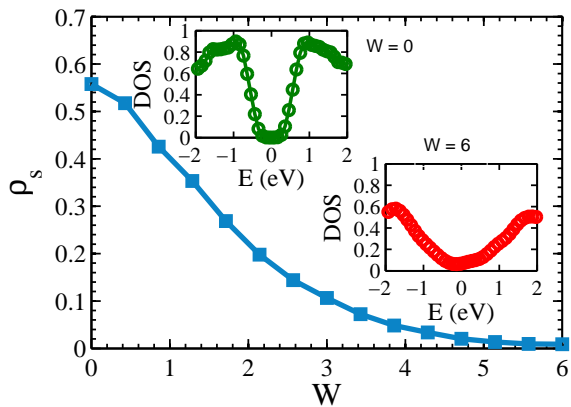, width=80mm}\vspace{-1.0em}
\caption{(Color online) Superfluid density  as a function of disorder.
  Insets  show   the  DOS  on   both  sides  of   the  superconducting
  transition.  Parameters used: $T  = 0$,  $H_x =  0.5$ and  $\alpha =
  0.8$.}
\label{stiffness_dos}
\end{center}
\end{figure}

Note  that the  translational invariance  is restored  via  a disorder
averaging so  that the correlation  function is only dependent  on the
separation $|r_{i}-r_{j}|$. As  shown in Fig.~\ref{stiffness_dos}, the
superfluid  density $\rho_s$  rapidly  declines with  the strength  of
disorder,  nearly  vanishing  for   $W\simeq  6$  and  signifying  the
proximity  of  a  superconductor  to insulator  transition  driven  by
disorder. Although  quantum phase fluctuations play a  crucial role in
the   superconductor-insulator  transition   in  the   large  disorder
regime~\cite{PhysRevB.65.014501,dubinature2007},  it   is  beyond  the
scope   of   the   current   approach  to   incorporate   such   phase
fluctuations. However,  from the  large suppression of  the superfluid
density  by  disorder,  we  can  clearly observe  the  destruction  of
superconductivity as  disorder strength increases.   The sharp decline
of stiffness with disorder  is understandable in the present situation
where  there  is  large  spatial fluctuation  of  the  superconducting
amplitude.  Since  the   stiffness  measures  rigidity  against  phase
fluctuations, and  in the mean-field  BdG theory the phase  is uniform
throughout,  under a  phase twist  at  the boundary  the system  would
accommodate steep adjustments of the phase change in regions where the
superconducting amplitude  is small.  This destroys  the overall phase
rigidity rapidly over the islands with vanishing gap thereby producing
the observed sensitivity to disorder.

\subsubsection{Finite Temperature Behaviour} 
According  to scanning squid  data~\cite{Bert2011}, superconductivity,
albeit   spatially    inhomogeneous,   appears   below    a   critical
temperature.  Also,  no temperature  dependence  of the  ferromagnetic
landscape has  been observed over  the measured temperature  range. On
the  other  hand,  the torque  magnetometry  measurement~\cite{Li2011}
reports  a coexistence of  ferromagnetism and  superconductivity below
120 mK with a superparamagnetic behavior, presumably from the magnetic
domains, persisting beyond  200 mK. We therefore undertake  a study of
the  system  at finite  temperatures  with  moderate  disorder. For  a
homogeneous  system the mean-field  superconducting transition  can be
determined  by the  condition  that the  mean superconducting  pairing
amplitude vanishes.   In an inhomogeneous situation,  such a criterion
is  no longer valid.  In practice,  the superconducting  transition is
determined  by the percolation  of the  superconducting regions  in an
insulating matrix,  whereas an useful indicator for  the transition is
the vanishing of the correlation function.
\begin{figure}[!ht]
\begin{center}
\epsfig{file=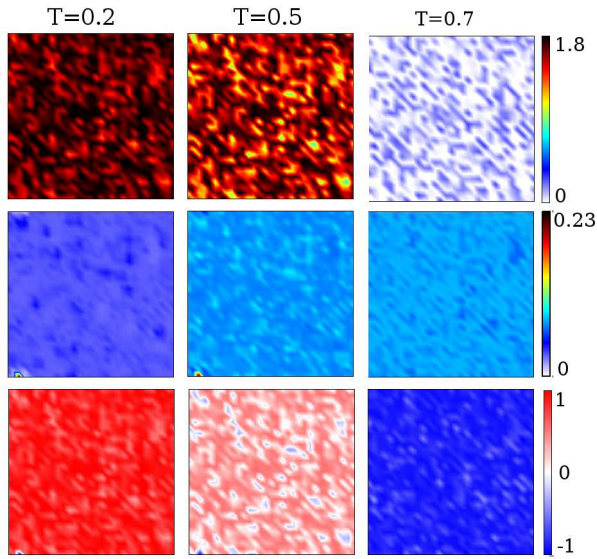, width=80mm}
\caption{(Color online) The spatial  distribution of the local pairing
  amplitude  $|\Delta(r_i)|$  (top   row)  and  magnetization  (middle
  row).  In  the  bottom  row,  red  and  blue  represent  regions  of
  superconductivity and magnetism  respectively. The three columns are
  for temperature $T =  0.2$, $T = 0.5$ and $T =  0.7$. The lowest row
  shows  coexistence  in combined  plots  of  magnetization and  local
  pairing amplitude. The parameters used are $H_x = 0.5$ and $\alpha =
  0.8$ at $W = 1$.}
\label{temp_evolution}
\end{center}
\end{figure}

\noindent  In  Fig.~\ref{temp_evolution},   we  show  the  temperature
evolution  of the  spatial  distribution of  local pairing  amplitude,
magnetization and  their coexistence  at different temperatures  for a
constant disorder strength ($W = 1$).  The top and middle row show, as
in Fig.~\ref{disorder_maps}, the  development of superconductivity and
magnetism  separately. With increasing  temperature, superconductivity
collapses    rapidly     nearly    vanishing    by     $T    =    0.7$
(Fig.~\ref{temp_evolution}, top  row, from left  to right), signifying
the    onset    of    superconducting   transition.     Concomitantly,
ferromagnetism  develops  (left  to  right,  middle row)  due  to  the
in-plane Zeeman  field as the temperature is  raised.  The coexistence
of superconductivity and ferromagnetism is  shown in the lowest row of
Fig.~\ref{temp_evolution}    as    temperature    rises.     Initially
superconductivity  dominates over  the entire  system  while magnetism
shows  up in  the  non-superconducting regions  as temperature  rises,
eventually setting  up a near-uniform ferromagnetic  state (with small
spatial fluctuations)  at higher temperatures ($T \simeq  0.7$) in the
entire area of the interface.
\begin{figure}[!ht]
\begin{center}
\epsfig{file=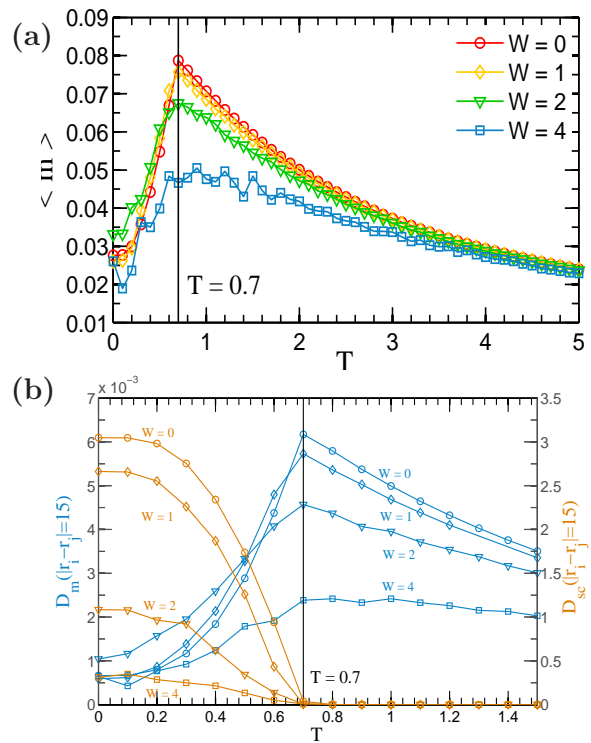, width=80mm}
\caption{(Color   online)  Temperature   variations  of   (a)  average
  magnetizations and (b)  correlation functions for different disorder
  strengths.   Beyond   $T   =   0.7$,   superconducting   correlation
  vanishes. The parameters used: $H_x = 0.5$, $\alpha = 0.8$.}
\label{temp_var}
\end{center}
\end{figure}

An  interesting  point to  note  is  that,  the average  magnetization
increases  with  temperature showing  a  peak near  $T  =  0.7$ as  in
Fig.~\ref{temp_var}(a).   This    observation   is   understood   from
Fig.~\ref{temp_var}(b), where  we plot the correlation  functions as a
function  of  temperature.  Beyond  $T  =  0.7$,  the  superconducting
correlation  is  nearly  absent,  ferromagnetism  peaks  (albeit  with
spatial fluctuations due  to disorder) and then drops  further on. The
tail  in  Fig.~\ref{temp_var}(a) is  extended  to higher  temperatures
depending on the Zeeman field  and disorder strength.  This picture is
quite similar to the torque magnetometry results mentioned above where
high  temperature magnetism is  found to  survive beyond  the putative
superconducting transition temperature.

\section{Discussions}
\label{discussions}
\begin{figure}[!ht]
\begin{center}
\epsfig{file=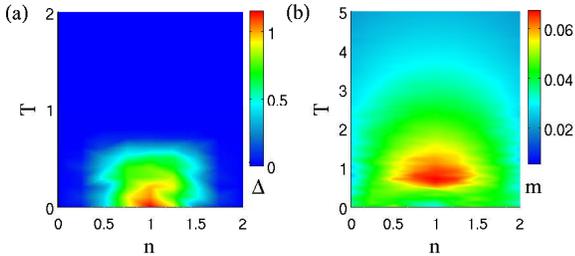, width=80mm}
\caption{(Color online) Plots of (a) average pairing amplitude and (b)
  magnetization  in  the  $n-T$  space  for  fixed  disorder  strength
  $W=2$. Parameters used: $H_x = 0.5$, $\alpha = 0.8$.}
\label{phase}
\end{center}
\end{figure}

Our calculations  provide a simple,  yet effective description  of the
superconductivity,  ferromagnetism and  their coexistence  observed at
the interface  at very low  temperatures.  We take  a phenomenological
model                                                               for
superconductivity~\cite{PhysRevLett.14.305,jourdan2003superconductivity}
in the presence  of local moments along with  Rashba SO interaction at
the  interface,  favoring  $p_x   \pm  ip_y$-wave  pairing  at  finite
momentum. The localized moments  interact with the itinerant electrons
via a  ferromagnetic exchange coupling acting like  an in-plane Zeeman
field. Our  calculations reveal that the non-magnetic  disorder play a
decisive  role  in  the   emergence  of  the  coexisting  phase  where
superconductivity   and    ferromagnetism   are   phase-separated   at
microscopic scale.  In the  homogeneous case, robust ferromagnetism is
absent and  superconductivity pervades the  two-dimensional interface.
Ferromagnetism  in  the  interface  is facilitated  by  the  disorder:
isolated  superconducting and  insulating regions  lead  to coexisting
superconductivity and magnetism.  However at strong disorder, both the
phases are  affected by large  local density fluctuations and  tend to
disappear.   For concreteness,  we  also study  the  behaviour of  the
system  at other fillings  as depicted  in Fig.~\ref{phase},  where we
plot the  mean pairing ampitude  and magnetization in the  $n-T$ plane
($n$ is  occupation number).  As  usual, superconductivity has  a dome
around half-filling  and a robust ferromagnetism  is established after
the superconducting phase is  degraded by disorder. However, the phase
diagram shows  a region of  coexistence of an inhomogenous  mixture of
superconductivity and ferromagnetism.

\noindent The general features  of magnetism and superconductivity are
reasonably well understood  in terms of a single  band model, although
one could envisage  a scenario where they occur  in different bands as
well. The essential  physics is disorder-driven and it  is possible to
explain the coexistence of  superconductivity and magnetism within the
framework of  a single band. Disorder-induced  ferromagnetism has also
been   observed   in   bulk   SrTiO$_3$   substrate   supporting   the
assertion~\cite{PhysRevX.2.021014,crandles:053908}  that  there  is  a
definite  connection between  the  presence of  impurity and  observed
ferromagnetism.

\begin{figure}[!ht]
\begin{center}
\epsfig{file=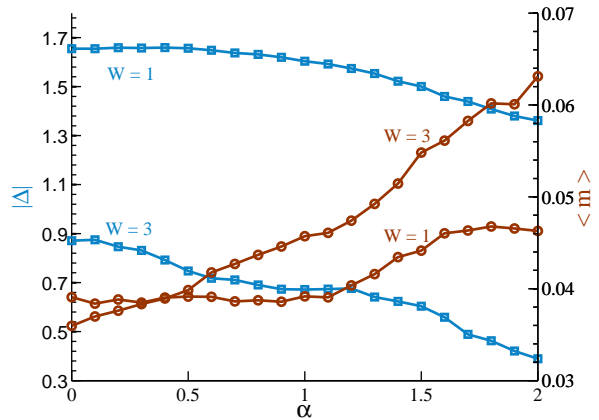, width=80mm}
\caption{(Color   online)  Variations  of   disorder-averaged  pairing
  amplitudes and magnetization with  Rashba coupling strength for $W =
  1$ and $W = 3$ with $H_x = 0.5$ and $T = 0$.}
\label{rashba_var}
\end{center}
\end{figure}

Rashba  spin-orbit interaction favors  an odd  pairing superconducting
state and a helical alignment of the electron spins without an overall
net moment.   However, the  presence of in-plane  exchange interaction
and intrinsic disorder are likely  to mitigate its effects and prevent
such a long range helical magnetic order. Therefore, phase segregation
is a  natural and  most likely outcome  as shown. Rashba  coupling, in
fact,  converts the  \textit{s}-wave superconductivity  into  a chiral
\textit{p}-wave  pairing  which  is  very  sensitive  to  non-magnetic
disorder. The presence  of the in-plane field and  the finite momentum
pairing then  lead to spin  precession at different rates  (the effect
being weak, as the Zeeman field is much weaker than spin-orbit scales)
for the  partners of  the pair depending  on the  spin-orbit coupling.
The dephasing coming from  slightly different spin precession rates of
the  pair  and  disorder   scattering  effectively  act  as  a  strong
pair-breaking    mechanism,    with    disorder    affecting    fairly
dramatically. As shown  in Fig.~\ref{rashba_var}, superconductivity is
weakly affected by  the spin-orbit coupling and while  it is degraded,
net  magnetic  moment  increases  concomitantly at  larger  spin-orbit
coupling.

The Oxygen vacancies at  the interface also have significant influence
on      the     magnetic      properties     of      the     interface
electrons~\cite{Schlom2011, PhysRevB.85.020407, 2013arXiv1311.1791M}. There is a
quenching of  magnetic moment  when the system  is annealed  in Oxygen
environment~\cite{2013arXiv1305.2226S}.  An  oxygen vacancy presumably
adds  two additional  electrons  which are,  most probably,  partially
localized  near  the  vacancy  due  to  Coulomb  correlation  and  the
reduction in  magnetic moment after  annealing is probably due  to the
reduction of Oxygen vacancies by molecular Oxygen.

Rashba  spin-orbit coupling  in presence  of an  applied perpendicular
Zeeman field leads  to the appearence of spinless  ${p_x + ip_y}$-wave
superconductivity  which   is  a  canonical   example  of  topological
superconductor   hosting   Majorana   bound   states   under   certain
conditions~\cite{PhysRevB.82.134521,PhysRevB.79.094504,PhysRevB.81.125318}.
Since the intrinsic Zeeman field  in the system is along the interface
plane and brings asymmetry in  the Fermi-surface, it is quite unlikely
to find any Majorana state here.  Moreover, the strong disorder is not
conducive          to           any          such          topological
excitations~\cite{PhysRevB.83.184520}.

\section{Conclusions} 
\label{conclusion}
In  the   foregoing,  we  presented  a  model   which  elucidates  the
coexistence   of   superconductivity   and   ferromagnetism   at   the
SrTiO$_3$/LaAlO$_3$  interface.  The  presence of  gate-tunable Rashba
coupling induces chiral $p_x  \pm ip_y$-wave superconductivity and the
asymmetric  Fermi-surface  due to  the  in-plane  Zeeman field  favors
pairing of electrons at  finite momentum. Spatial inhomogeneity at the
interface plays a key role in microscopic separation of ferromagnetism
and superconductivity  and hence the  observed coexistence of  the two
phases.  With large  disorder, the  electronic system  segregates into
superconducting patches  and in  the insulating regions  local moments
order and  form ferromagnetic puddles.  Our scenario  accounts for the
scanning      squid~\cite{Bert2011,     doi:10.1021/nl301451e}     and
magneto-resistance~\cite{Wang2011}    measurements    which    suggest
electronic phase separation in the Ti $d$-bands at the interface.

Similar phenomena of the formation of 2DEG, exhibiting low-temperature
superconductivity  and  ferromagnetism,  have  also been  reported  in
epitaxially                  grown                 GdTiO$_3$/SrTiO$_3$
interface~\cite{2012PhRvX...2b1014M}.  The  present  work  provides  a
unique platform  for studying novel  interfacial superconductivity and
other interesting properties in such heterostructures.

\section*{Acknowledgements}
The authors thank S.   S.  Mandal, G.  Baskaran and Titus Neupert
for  useful discussions.   NM acknowledges  MHRD, India  for financial
assistance  and  AT acknowledges  CSIR,  India  for financial  support
through a joint project.


\end{document}